\documentclass[12pt,a4paper,oneside]{article}
\usepackage[dvips]{graphicx}
\newcommand{\nucl}[2]{{}^{#1}\mathrm{#2}}
\begin{document}

\begin{center}
{\large
 {\bf Spin-Isospin Excitations and Muon Capture by Nuclei}

 V.\,A.\ Kuz'min${}^{a)}$ and T.\,V.\ Tetereva${}^{b)}$ \\

 }

 {\it
   ${}^{a)}$ Joint Institute for Nuclear Research,
        Dubna, Moscow region, 141980, Russia  \\
   ${}^{b)}$ Skobeltsyn Institute of Nuclear Physics, Lomonosov
        Moscow State University, Moscow, 119992, Russia

 }
\end{center}

\begin{abstract}
\noindent
 By analyzing the energy-weighted moments of the strength function
 calculated in RPA and beyond it is shown that
 the explanation of the effect of missing strength of Gamow-Teller
 transitions requires that residual interaction
 produce high-excited $1^{+}$ particle-hole collective states.
 The example of this interaction is presented.
 The manifestations of spin-isospin nuclear response in
 nuclear muon capture are discussed.
\end{abstract}

\section*{Introduction}

 To discuss the problem of missing strength
 of Gamow-Teller (GT) transitions, one needs to study the
 distribution of transition strength over the excitation energy.
 A convenient tool for that purpose is the strength function of
 GT transitions
 \begin{equation}
 \label{eq:str_function}
  b^{\pm}(E) = \sum_{\alpha} \sum_{\mu = -1}^{1}
   \Bigl \vert \langle \alpha \biggl \vert
   \sum_{n=1}^{A} \sigma_{\mu}(n)\; t^{\pm}(n)
   \biggr \vert \mathrm{g.s} \rangle \Bigr \vert^{2}
   \delta\left(E-E_{\alpha}\right) ,
 \end{equation}
 where $E_{\alpha}$ is the energy of the $\vert \alpha \rangle$ state
 reckoned from the ground state of a target nucleus,
 $\vert \mathrm{g.s} \rangle$.

 Usually, it is assumed that the effect of missing strength
 can be explained (reproduced) by including 2p--2h admixtures into
 the wave functions of nuclear states involved.
 In order to check this assumption, we consider
 the energy-weighted moments of the GT strength function
\begin{equation}
\label{eq:moments}
  S^{\pm}_{k} = \int_{0}^{\infty} \, E^{k} b^{\pm}(E) \, dE
\end{equation}
 for positive integer $k$.
 In the first section, we calculate the moments in
 the random phase approximation,
 the second random phase approximation
 and within the fragmentation problem.
 In the second section, we argue that the explanation
 of missing strength requires that the particle-hole
 residual interaction has a specific feature,
 it should intensively mix the $\Delta N = 0$ and
 $\Delta N \geq 2$ particle-hole configurations.
 The example of such an interaction is presented and
 the strength function of $\sigma t^{-}$ transitions
 in $\nucl{90}{Zr}$ is demonstrated.
 In the third section, the calculations of total
 muon capture rates by complex nuclei are discussed.
 The fourth section contains the analysis of spin-isospin transitions
 in $A=28$ nuclei observed in muon capture, and in $(e,e^{\prime})$
 and $(p,n)$ reactions.
 The main results are collected in Conclusion.

\section{Energy-weighted moments of strength function}

 In this section we give in a compact form a summary of the basic
 equations for energy-weighted moments calculated
 in the random phase approximation (RPA),
 the second random phase approximation (SRPA) and
 within the fragmentation problem.

\subsection{Random Phase Approximation}

 The formulae are presented for double-magic nuclei;
 the extension to the open shell nuclei is straightforward ---
 one should replace the creation operators of particle-hole states
 by two-quasiparticle ones and the Hartree-Fock ground state
 by the state of quasiparticle vacuum.
 We use the labels $h$, $h^{\prime}$ for the
 occupied single-particle states and the labels $p$, $p^{\prime}$
 for vacant ones.
 The $\Phi_0$ is a Slater determinant consisting of occupied states
 only.
 For a magic number system $\Phi_0$ is the nondegenerated ground
 states in the model of independent particles.
 The phonon creation and destruction operators are defined by
\begin{equation}
\label{eq:rpa_phonon}
\matrix{
 \displaystyle
   O^{\dag}_{\alpha}  =
   \sum_{p,h} \left ( \psi^{\alpha}_{p,h} a^{\dag}_{p} a_{h}
         - \phi^{\alpha}_{p,h} a^{\dag}_{h} a_{p} \right ) , \cr
 \noalign{\smallskip}
 \displaystyle
   O_{\alpha}  =
   \sum_{p,h} \left ( \bar{\psi}^{\alpha}_{p,h} a^{\dag}_{h} a_{p}
                    - \bar{\phi}^{\alpha}_{p,h} a^{\dag}_{p} a_{h} \right ) . \cr
}
\end{equation}
 The phonon amplitudes $\psi^{\alpha}_{p,h}$ and $\phi^{\alpha}_{p,h}$
 are determined by the homogeneous system of linear equations:
\begin{equation}
\label{eq:eq_motion}
 \matrix{
 \displaystyle
  \langle \Phi_{0} \vert
    \bigl [ [ O_{\alpha}, H ], a^{\dag}_{p} a_{h} \bigr ]
     \vert \Phi_{0} \rangle
     &  = &
 \displaystyle
  E_{\alpha}
  \langle \Phi_0 \vert
     \bigl [ O_{\alpha}, a^{\dag}_{p} a_{h} \bigr ]
      \vert \Phi_{0} \rangle , \cr
 \noalign{\smallskip}
 \displaystyle
  \langle \Phi_{0} \vert
    \bigl [ [ O_{\alpha}, H ], a^{\dag}_{h} a_{p} \bigr ]
     \vert \Phi_{0} \rangle
     & = &
 \displaystyle
  E_{\alpha}
  \langle \Phi_0 \vert
     \bigl [ O_{\alpha}, a^{\dag}_{h} a_{p} \bigr ]
      \vert \Phi_{0} \rangle . \cr
 }
\end{equation}
 These equations can be obtained by either the
 ``equation-of-motion'' method of Rowe \cite{Rowe68}
 or a time-dependent variational principle \cite{LaneMart80}.
 After the complex conjugate one has
\begin{equation}
 \label{eq:RPA_eigenproblem}
  \pmatrix{ A & D \cr D^{\ast} & A^{\ast} \cr }
  \pmatrix{ \psi^{\alpha} \cr \phi^{\alpha} \cr } =
  E_{\alpha} \pmatrix{ \psi^{\alpha} \cr -\phi^{\alpha} } ,
\end{equation}
 where $A$ and $D$ are the square matrices with the elements
\begin{equation}
\label{eq:rpa_matrix}
\matrix{
 \displaystyle
 A_{p h,p^{\prime} h^{\prime}} &
 \displaystyle
 =
 \langle \Phi_{0} \vert a^{\dag}_{h} a_{p} H
    a^{\dag}_{p^{\prime}} a_{h^{\prime}} \vert \Phi_{0} \rangle
   - \langle \Phi_{0} \vert H \vert \Phi_{0} \rangle
 \delta_{p, p^{\prime}} \delta_{h, h^{\prime}} \hfill \cr
\noalign{\smallskip}
 \displaystyle
 D_{p h, p^{\prime} h^{\prime}} &
 \displaystyle
 =
 \langle \Phi_{0} \vert a^{\dag}_{h} a_{p}
   a^{\dag}_{h^{\prime}} a_{p^{\prime}} H
    \vert \Phi_{0} \rangle . \hfill \cr
 }
\end{equation}
 As the Hamiltonian $H$ of the system is Hermitian, $A$ and $D$
 are Hermitian symmetric matrices.
 From the theory of regular pencils of matrices \cite{Gantmacher}
 it follows that for the positive-definite matrix
 $ \pmatrix{ A & D \cr D^{\ast} & A^{\ast} \cr }$
 (in this case $\Phi_{0}$ is stable against particle-hole
 excitations) the system (\ref{eq:RPA_eigenproblem}) has
 only nonzero eigenvalues $E_{\alpha}$.
 The corresponding nontrivial solutions can be normalized by
\begin{equation}
  \langle \Phi_{0} \vert
   \bigl [ O_{\alpha}, O^{\dag}_{\beta} \bigr ]
    \vert \Phi_0 \rangle =
  \mathrm{sign}(E_{\alpha}) \, \delta_{\alpha,\beta} ;
\end{equation}
 Positive $E_{\alpha}$'s
 are considered as approximate excitation energies
 of the system.

 For any one-body transition operator $R$ the matrix elements between
 the ground $\vert 0 \rangle$ and excited $\vert \alpha \rangle$ states,
 calculated in the RPA, are defined by
\begin{equation}
 \label{eq:RPA_mat_el}
 R_{\alpha} \equiv
 \langle \alpha \vert R \vert 0 \rangle_{\mathrm{RPA}} \equiv
  \langle \Phi_{0} \vert
   \bigl [ O_{\alpha}, R \bigr ] \vert \Phi_{0} \rangle .
\end{equation}
 The normalization condition for a one-phonon state with positive
 energy $E_{\alpha}$ can be obtained from this equation
\begin{displaymath}
 \langle \alpha \vert O^{\dag}_{\beta} \vert 0 \rangle_{\mathrm{RPA}} =
  \langle \Phi_{0} \vert
   \bigl [ O_{\alpha}, O^{\dag}_{\beta} \bigr ] \vert \Phi_{0} \rangle
   = \delta_{\alpha,\beta} .
\end{displaymath}
 The spectral properties of the eigenvalue
 problem (\ref{eq:RPA_eigenproblem}) allow one to write down
\begin{equation}
\label{eq:RPA_spectral}
 \sum_{\alpha} \mathrm{sign}(E_{\alpha}) \, E^{k}_{\alpha}
 \pmatrix{ \psi^{\alpha} \cr \phi^{\alpha} \cr }
 \pmatrix{ {\psi^{\alpha}}^{\dag} & {\phi^{\alpha}}^{\dag} \cr } =
 \pmatrix{ A & D \cr - D^{\ast} & - A^{\ast} \cr }^{\! k}
 \pmatrix{ \hat{1} & \hat{0} \cr \hat{0} & - \hat{1} \cr } ,
\end{equation}
 for any integer $k$.
 The square matrices $\hat{1}$ and $\hat{0}$ are unit and zero ones.
 The energy-weighted moments for the transition strength of
 any one-body transition operator $R$ are obtained from
 (\ref{eq:RPA_spectral})
\begin{equation}
\label{eq:RPA_miments}
 \sum_{\alpha: \, E_{\alpha} > 0}
 E_{\alpha}^{k} \, \bigl ( \vert R_{\alpha} \vert^2
  - (-1)^{k} \vert R^{\dag}_{\alpha} \vert^2 \bigr ) =
 \pmatrix{ R^{\dag} & \widetilde{R^{+}} \cr}
 \pmatrix{ A & D \cr - D^{\ast} & - A^{\ast} \cr }^{k}
 \pmatrix{ R \cr - {R^{+}}^{\ast} \cr }.
\end{equation}
 The components of the vectors $\pmatrix{R}$ and $\pmatrix{R^{+}}$
 are the matrix elements
 $ \langle \Phi_{0} \vert a^{\dag}_{h} a_{p} R \vert \Phi_{0} \rangle $
 and
 $ \langle \Phi_{0} \vert a^{\dag}_{h} a_{p} R^{\dag} \vert \Phi_{0} \rangle $,
 respectively.

\subsection{Second Random Phase Approximation}

 Within the Second Random Phase Approximation (SRPA) \cite{DNSW90},
 the definition of the phonon operators (\ref{eq:rpa_phonon})
 is extended by including the creation and destruction operators
 of the two-particle--two-hole excitations
\begin{displaymath}
  \mathcal{O}^{\dag}_{\alpha} =  \sum_{p,h} \, \left (
    \psi^{\alpha}_{p,h} a^{\dag}_{p} a_{h} -
    \phi^{\alpha}_{p,h} a^{\dag}_{h} a_{p} \right )
  +  \sum_{p < p^{\prime}, h < h^{\prime} } \, \left (
    \psi^{\alpha}_{p p^{\prime}, h h^{\prime}}
       a^{\dag}_{p} a^{\dag}_{p^{\prime}} a_{h^{\prime}} a_{h} -
    \phi^{\alpha}_{p p^{\prime}, h h^{\prime}}
       a^{\dag}_{h} a^{\dag}_{h^{\prime}} a_{p^{\prime}} a_{p}
    \right ) .
\end{displaymath}
 Phonon amplitudes are determined by (\ref{eq:eq_motion}) after
 replacing the RPA phonons by the SRPA ones \cite{DNSW90}.
 From (\ref{eq:eq_motion}) one obtains
\begin{displaymath}
 \pmatrix{
     \mathcal{A} & \mathcal{D} \cr
     \mathcal{D}^{\ast} & \mathcal{A}^{\ast} \cr }
  \pmatrix{ \chi^{\alpha} \cr \rho^{\alpha} \cr }
  =
  \mathcal{E}_{\alpha}
  \pmatrix{ \chi^{\alpha} \cr - \rho^{\alpha} \cr }
\end{displaymath}
 where
\begin{displaymath}
\matrix{
 \displaystyle
 \mathcal{A} =
  \pmatrix{ A_{p h, p^{\prime} h^{\prime}} \hfil &
     A_{p h, p^{\prime}_{1} p^{\prime}_{2}
           h^{\prime}_{1} h^{\prime}_{2} } \hfil \cr
     A_{p_{1} p_{2} h_{1} h_{2}, p^{\prime} h^{\prime} } \hfil &
     A_{p_{1} p_{2} h_{1} h_{2},
    p^{\prime}_{1} p^{\prime}_{2} h^{\prime}_{1} h^{\prime}_{2} } \hfil \cr } ,\cr
 \noalign{\smallskip}
 \displaystyle
 \mathcal{D} =
 \pmatrix{ D_{p h , p^{\prime} h^{\prime}} \hfil &
     D_{p h, p^{\prime}_{1} p^{\prime}_{2}
           h^{\prime}_{1} h^{\prime}_{2} } \hfil \cr
     D_{p_{1} p_{2} h_{1} h_{2}, p^{\prime} h^{\prime} } \hfil &
     D_{p_{1} p_{2} h_{1} h_{2},
    p^{\prime}_{1} p^{\prime}_{2} h^{\prime}_{1} h^{\prime}_{2} } \hfil \cr } \cr
 }
\end{displaymath}
and
\begin{displaymath}
  \chi^{\alpha} =
  \pmatrix{ \psi^{\alpha}_{p h} \hfil \cr
            \psi^{\alpha}_{p_{1} p_{2} h_{1} h_{2}} \hfil \cr } ,
 \qquad
  \rho^{\alpha} =
  \pmatrix{ \phi^{\alpha}_{p h} \hfil \cr
            \phi^{\alpha}_{p_{1} p_{2} h_{1} h_{2}} \hfil \cr } .
\end{displaymath}
 The algebraic structure of the equations remains the same as in the RPA.
 This is true regarding the formal definition of the matrix elements
 of the transition operator $R$,
 $ \mathcal{R}_{\alpha} \equiv
 \langle \alpha \vert R \vert 0 \rangle_{\mathrm{SRPA}} \equiv
  \langle \Phi_{0} \vert
   \bigl [ \mathcal{O}_{\alpha}, R \bigr ] \vert \Phi_{0} \rangle $.
 The distinction is the appearance of SRPA phonons instead
 of RPA phonons.
 The only nonvanishing matrix elements of any one-body transition
 operator are those between $\Phi_{0}$ and the 1p--1h excited states.
 As a consequence, the zero energy-weighted moment reduces to
\begin{displaymath}
 S^{-}_{0} - S^{+}_{0} =
 \sum_{p,h} \left (
   \bigl \vert \langle \Phi_{0} \vert a^{\dag}_{h} a_{p} R
        \vert \Phi_{0} \rangle \bigr \vert^{2}
 - \bigl \vert \langle \Phi_{0} \vert a^{\dag}_{h} a_{p} R^{\dag}
        \vert \Phi_{0} \rangle \bigr \vert^{2}
 \right )
\end{displaymath}
 (the length of the vector $(R)$ minus the length of $(R^{+})$).
 This difference is the same in both the approximations,
 because it is determined by the space of particle-hole
 excitations only.
 For the Gamow-Teller transitions
 ($ R_{\mu} = \sigma_{\mu} t^{-} $
 and $ R^{\dag}_{\mu} = (-1)^{\mu} \sigma_{-\mu} t^{+} $) the zero
 energy-weighted moment reduces to $3(N-Z)$ which is the value
 of the Ikeda sum rule.

 The first energy-weighted moment calculated in the SRPA is
\begin{displaymath}
 S^{-}_{1} + S^{+}_{1} =
 \sum_{\alpha: \mathcal{E}_{\alpha} > 0}
 \mathcal{E}_{\alpha} \, \left (
   \vert \mathcal{R}_{\alpha} \vert^{2} + \vert \mathcal{R}^{\dag}_{\alpha} \vert^{2}
 \right ) =
 \pmatrix{ R^{\dag} & \widetilde{R^{+}} \cr }
 \pmatrix{ A & D \cr -D^{\ast} & -A^{\ast} \cr }
 \pmatrix{ R \cr -{R^{+}}^{\ast} \cr } .
\end{displaymath}
 It coincides with the one obtained in the RPA.

\subsection{Fragmentation problem}

 A similar situation appears within the fragmentation
 problem \cite{BM,VS1983},
 where the ground state  $\vert 0\rangle$ is assumed to
 be presented by some model wave function, and one studies
 the influence of the enlarging of the space excited states
 on the distribution of transition strength.
 The wave function of the excited state can be decomposed
 into some basis
\begin{displaymath}
 \Psi_{\alpha} = \sum_{m} c_{\alpha,m} \psi_{m} +
   \sum_{n} \tilde{c}_{\alpha,n} \tilde{\psi}_{n} .
\end{displaymath}
 In this expression the set of basis vectors has been divided into
 two parts, according to the values of the matrix elements of the
 transition operator $R$
\begin{displaymath}
  \langle \psi_{m} \vert R \vert 0 \rangle \neq 0 \ , \qquad
  \langle \tilde{\psi}_{n} \vert R \vert 0 \rangle = 0 \ .
\end{displaymath}
 For example, if one used the Hartree-Fock ground state
 and the space of excited
 states spanned by the 1p--1h and 2p--2h basis vectors,
 the particle-hole components would belong to the first set
 and the two-particle--two-hole vectors would be ones of
 the second group for every one-body transition operator.

 It was proved in \cite{KuzFrag} that in this problem the zero and first
 energy-weighted moments ($S_{0}$ and $S_{1}$) are determined
 by the $\psi$-subspace of simple excited states only,
 and do not depend on the interaction between $\psi$- and
 more complicated $\tilde{\psi}$-states and on the
 interactions acting inside the $\tilde{\psi}$-subspace alone.
 This is a direct consequence of freezing the ground state.

\section{Missing strength and nuclear residual interaction}

\subsection{Analysis of energy-weighted moments}

 The zero and first energy-weighted moments are conserved when
 going from RPA to SRPA
\begin{equation}
\label{eq:srpa_zero}
 S^{-}_{0} - S^{+}_{0} \bigr \vert_{\mathrm{RPA}} =
 S^{-}_{0} - S^{+}_{0} \bigr \vert_{\mathrm{SRPA}}
\end{equation}
 and
\begin{equation}
\label{eq:srpa_one}
 S^{-}_{1} + S^{+}_{1} \bigr \vert_{\mathrm{RPA}} =
 S^{-}_{1} + S^{+}_{1} \bigr \vert_{\mathrm{SRPA}} .
\end{equation}
 Within the fragmentation problem it was
 shown \cite{KuzFrag} that $S_{0}$ and $S_{1}$ are
 determined by the 1p--1h states only and do not depend on
 the interaction between the 1p--1h states and 2p--2h states or
 more complex states and on any interaction in the
 2p--2h subspace,
\begin{equation}
\label{eq:frag_zero}
   S^{\pm}_{0} \bigr \vert_{\mathrm{RPA}} =
   S^{\pm}_{0} \bigr \vert_{\mathrm{fragmentation}}
\end{equation}
 and
\begin{equation}
\label{eq:frag_one}
   S^{\pm}_{1} \bigr \vert_{\mathrm{RPA}} =
   S^{\pm}_{1} \bigr \vert_{\mathrm{fragmentation}} .
\end{equation}
 It is important to stress that equations
 (\ref{eq:srpa_zero} -- \ref{eq:frag_one}) follow
 from the properties of algebraic equations to be solved.
 So they should be valid for results of {\em any}
 calculation.

 It is known that the interaction between the 1p--1h and 2p--2h
 states is responsible for the width of the giant resonance and
 causes some redistribution of the transition strength over
 the excitations energy \cite{Bert83,VS1983}.
 The conservation of zero and first moments has, however, a severe
 consequence for the problem of missing strength.
 We discuss it in the framework of the fragmentation problem,
 in which the moments $S_{0}^{\pm}$ and $S_{1}^{\pm}$ are
 separately conserved.
 As total transition strength and average
 excitation energy, $S^{-}_{1}/S^{-}_{0}$, are simultaneously conserved,
 we face the following situation.
 If an interaction between 1p--1h states and more complex
 states moves a large fraction of the strength of the giant resonance
 to higher energies, then some strength has to be shifted
 into the low energy region in order to keep the ratio
 $S^{-}_{1}/S^{-}_{0}$ constant.
 As a result, the strength distribution in the giant resonance region
 and below it would change completely.
 Such an effect has been found by shell model calculations
 of the GT strength function \cite{Mat83}, where
 the excited state space was enlarged by including
 the 2p--2h configurations,
 the ground state was left untouched, and
 the described effect was exactly observed.

 The authors of \cite{DKSW86} calculated the GT strength function
 for $\nucl{48}{Ca}$, $\nucl{90}{Zr}$ and $\nucl{208}{Pb}$ in both
 the RPA and SRPA.
 The realistic two-body forces based on a Brueckner G-matrix
 were used as a nuclear residual interaction.
 According to the results of \cite{DKSW86}, a large fraction of the
 total GT strength was shifted towards higher excitation energies
 due to the interaction of the 1p--1h states with 2p--2h ones.
 Simultaneously, the strength in the giant resonance
 region and below it was considerably reduced.
 The energy of the GT resonance does not decrease and
 in $\nucl{208}{Pb}$ considerably increases.
 The subsequent calculations \cite{DO1987} show that in
 the $\nucl{90}{Zr}$ nucleus the average energies of GT transitions
 and spin-dipole $Y_{1} \sigma t^{-}$ transitions
 increased by $6\,\mathrm{MeV}$ in the SRPA.

 Now we consider equations (\ref{eq:srpa_zero}) and
 (\ref{eq:srpa_one}).
 Due to the large neutron excess in heavy nuclei the $\sigma t^{-}$
 strength, measured in $(p,n)$ reactions, is much stronger than
 the $\sigma t^{+}$ strength related to $(n,p)$ reactions.
 Therefore, $S_{0}^{+}$ is only a small fraction of $S_{0}^{-}$ and
 one can assume without making too big an error that
 $S_{0}^{-}$ and $S_{1}^{-}$ do practically not change when going
 from the RPA to SRPA.
 Then the arguments presented above for the fragmentation
 problem are applicable and a contradiction appears between
 the results of \cite{DKSW86,DO1987} and the
 conservation of moments $S_{0}^{-}$ and $S_{1}^{-}$.
 By this reason we cannot consider the results of
 \cite{DKSW86,DO1987} as correct ones.

\subsection{Missing strength and properties of residual interaction}

 The discussion leads to the conclusion that the transfer of the
 GT transition strength to higher excitation energies, which
 is implied by the effect of missing strength, originates rather
 from the interaction between particle--hole configurations only
 than from the influence of 2p--2h and more complex configurations.
 Large GT matrix elements exist between single particle states
 with identical radial and orbital quantum numbers.
 The corresponding transitions have relatively low energy.
 Therefore, residual interaction should mix two-quasiparticle states,
 those quasiparticles are from the same major shell
 ($\Delta N = 0$) with two-quasiparticle states containing
 quasiparticles from different major shells ($\Delta N \geq 2$).
 In others words, it follows from the very existence of the effect
 of missing GT strength that residual forces between nucleons must have
 a characteristic property: strong interaction
 between $\Delta N = 0$ and $\Delta N \geq 2$ two-quasiparticle
 (or particle-hole) states exists.

\begin{figure}
 \includegraphics[width=0.8\linewidth]{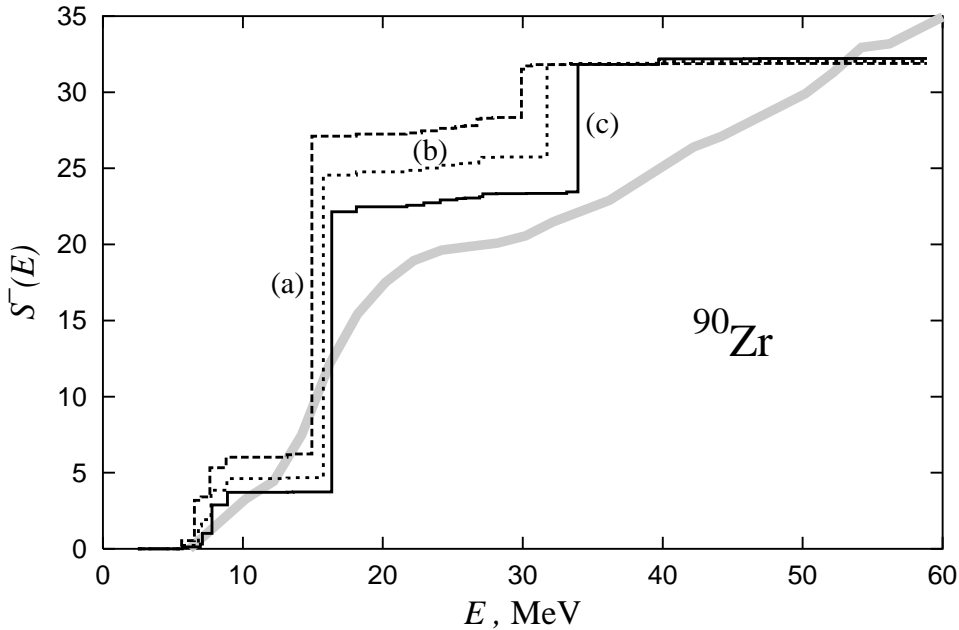}
\caption{%
 Distribution of $\sigma t^{-}$ strength in $\nucl{90}{Zr}$.
 Thick gray line is experimental distribution of the GT strength
 \protect\cite{Wakasa1997}.
 Lines labeled by (a), (b) and (c) present the results of
 calculations with different values of $\kappa^{01}_{1}$.}
\label{fig:zr90gt}
\end{figure}

 An example of such a residual interaction is given by the
 phenomenological separable residual interaction
 discussed in detail in \cite{VdoSol1983},
\begin{equation}
\label{eq:residual}
  H_\mathrm{res} = - \frac{1}{2}  \sum_{L,J,M} \sum_{q \neq q^{\prime}}
    \left ( \kappa^{LJ}_0 +
    \kappa^{LJ}_1 (\vec{\tau}_q \cdot \vec{\tau}_{q^{\prime}} )\right )
        Q^{\dag}_{LJM}(q)\, Q_{LJM}(q^{\prime}) .
\end{equation}
 Here $\vec{\tau}$ are isospin Pauli matrices, and
\begin{displaymath}
 Q_{LJM}(q) =
 i^{L} f_{LJ}(r_{q}) \, [Y_{L}(\hat{r}_{q}) \times \sigma_{q}]_{JM}
\end{displaymath}
 is the one-body spin-multipole operator.
 We have shown here the spin-multipole part of residual interaction,
 the expression for a multipole one can be found in \cite{VdoSol1983}.
 The interaction with the radial form factor
 \begin{equation}
 \label{eq:dUdr}
    f_{LJ}(r) = \frac{d U(r)}{d r},
 \end{equation}
 (here $U(r)$ is central part of Saxon-Woods potential)
 has a surface character and may mix $\Delta N = 0$ and $\Delta N \geq 2$
 particle-hole excitations.

 The results of calculation of the GT strength function in $\nucl{90}{Zr}$
 together with the experimental data of \cite{Wakasa1997}, where
 $(93 \pm 5)\%$ of the sum rule value $3(N-Z) = 30$ was observed,
 are presented in Fig.\,\ref{fig:zr90gt}.
 The GT strength function is shown as the running sum
 \begin{displaymath}
   S^{-}(E) = \int_{0}^{E} \, b^{-}(E^{\prime}) \, d E^{\prime} .
 \end{displaymath}
 The lines marked by (a), (b) and (c) are
 the theoretical strength functions
 calculated with $\kappa^{01}_{1} = -0.23 / A$,
 $\kappa^{01}_{1} = -0.43 / A$ and $\kappa^{01}_{1} = -0.63 / A$,
 respectively.
 It is easily seen in Fig.\,\ref{fig:zr90gt} that
 new high-lying collective $1^{+}$ states
 appear.
 When $\left\vert \kappa^{01}_{1} \right\vert$ grows the energy
 of the collective state and its $B(GT)$ increase.
 For others spherical nuclei theoretical GT strength functions
 have been calculated and compared with the experimental ones
 in \cite{examples}.
 All examples confirm that for explanation of
 the effect of missing (quenching) of the GT strength
 the specific feature of nuclear residual interaction is
 required: there should be a strong
 mixing among the $\Delta N = 0$ and $\Delta N \geq 2$ particle-hole
 configurations.
 The appearance of this specific feature of residual interaction should
 be checked in other charge-exchange processes
 related to spin-isospin transitions.

\section{Total rates of muon capture}

 The spin-isospin transitions are observed in
 two weak interaction processes: beta-decay and muon capture.
 Only a small fraction of the whole GT transition strength can usually
 be observed in beta decay because the limited energy release and
 variations in the GT strength function at low excitation
 energy affect noticeably theoretical $ft$-values.
 This limitation is removed in the reaction of ordinary muon capture
 (OMC)
\begin{equation}
\label{eq:OMC}
 \mu^{-} + A(Z,N) \rightarrow \nu_{\mu} + B(Z-1,N+1),
\end{equation}
 where states in a wide excitation energy range can be populated
 due to a relatively large muon mass,
 $ m_{\mu} \approx 105\,\mathrm{MeV}$.
 The phase space factor of partial OMC rate $\Lambda_{f}$ depends
 on the square of neutrino energy
 \begin{displaymath}
  E_{\nu} = \left ( m_{\mu} - \vert \epsilon_{1S} \vert + M_{A}
            - M_{B} - E_{f} \right )
  \left ( 1 - { { m_{\mu} - \vert \epsilon_{1S} \vert + M_{A}
      - M_{B} - E_{f} } \over { 2 ( m_{\mu} + M_{A} ) } } \right ),
 \end{displaymath}
 where $E_{f}$ is the excitation energy in the nucleus $B$,
 $M_{A}$ and $M_{B}$ are the masses of initial and final nuclei,
 $\epsilon_{1S}$ is the muon binding energy in the muonic atom.
 For example, for $0^{+} \rightarrow 1^{+}$ the rate of muon capture
\begin{equation}
\label{eq:partial_rate}
 \matrix{
 \Lambda_{f} \sim & \displaystyle
  E_{\nu}^{2} \, g_{A}^{2}
 \biggl \langle 1^{+}_{f} \biggl \Vert
  \sum_{q} j_0(E_{\nu} r_{q}) \, \sigma_{q}
    \, t^{+}_{q} \biggr \Vert 0^{+}_{\mathrm{g.s.}} \biggr \rangle^{2} \hfill \cr
 \noalign{\smallskip}
 {} & \displaystyle \times
  \biggl \{ 1 + {2 \over 3} \bigl [
  1 + 4 { g_{V} + g_{M} \over g_{A} }
   - {g_{P} \over g_{A} }
    \bigr ] { E_{\nu} \over 2 M_{p} }
   + { 1 \over 3 } \bigl (
     {g_{P} \over g_{A} }
     \, {E_{\nu} \over 2 M_{p} } \bigr )^{2}
 + \ldots \biggr \} , \hfill \cr
 }
\end{equation}
 here $M_{p}$ is the proton mass, and
 $g_{V}$, $g_{A}$, $g_{M}$ and $g_{P}$ are vector, axial-vector,
 weak-magnetic, and pseudoscalar couplings of weak
 nucleon current \cite{BKE1978}.

 The energy released in the muon capture is comparable
 with muon mass; therefore,
 the relative error in theoretical $\Lambda_{f}$ caused by
 uncertainties in the theoretical strength function for
 the corresponding nuclear transitions is rather small
 \begin{displaymath}
  {\Delta \Lambda_{f} \over \Lambda_{f} } \approx
  { \Delta E_{f} \over E_{\nu} } \ll { \Delta E_{f} \over E_{f} }
 \end{displaymath}
 in comparison with the ones in beta-decay,
 and one can hope that calculation in the RPA, which reproduces well
 the general features of strength functions of the corresponding transitions,
 will give a reasonable description of the OMC rates.
 The total muon capture rate is calculated as the sum of the rates for
 all possible nuclear transitions
\begin{displaymath}
 \Lambda_{\mathrm{tot}} = \sum_{f} \Lambda_{f} .
\end{displaymath}
\begin{figure}
 \includegraphics[width=0.8\linewidth]{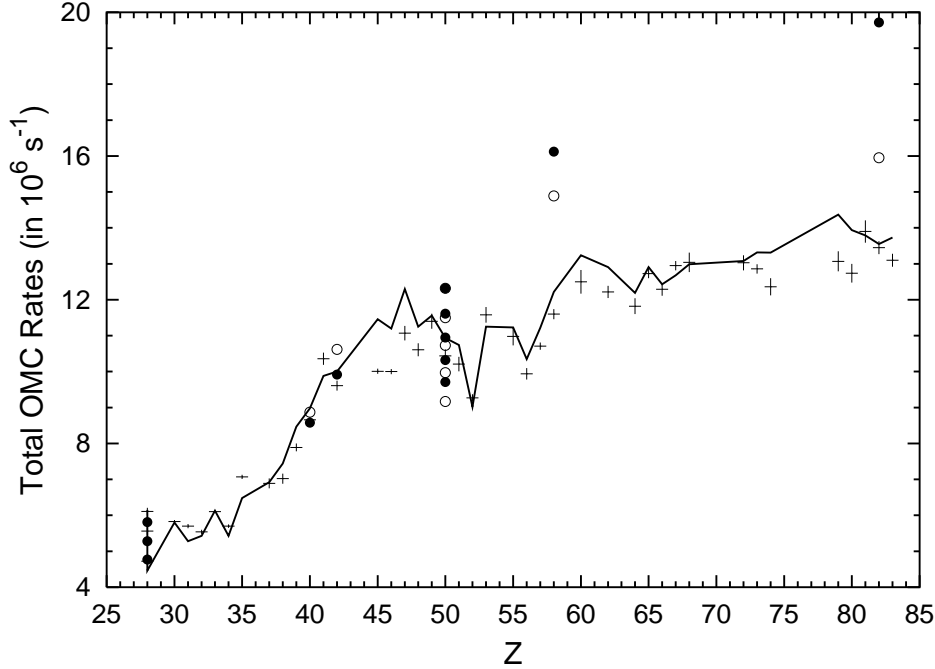}
 \caption{Total rates of OMC on complex nuclei.
  Open and closes circles are the calculation results;
  the crosses are the experimental data.
  See the text for details. }
 \label{fig:omc_tot}
\end{figure}
 The total OMC rates for several even-even spherical nuclei are
 calculated and discussed in detail in \cite{EKTJO}.
 Here we shortly present the main results.
 The $\Lambda_{\mathrm{tot}}$ are displayed in Fig.\,\ref{fig:omc_tot}.
 The results of calculations, in which the separable residual interaction
 (\ref{eq:residual}) with radial form factor (\ref{eq:dUdr}) has been used,
 are shown by the closed circles.
 For comparison the OMC rates were calculated with the radial
 form factor of residual interaction
 \begin{equation}
 \label{eq:r_L}
   f_{LJ}(r) = r^{L} .
 \end{equation}
 In Fig.\,\ref{fig:omc_tot}, the corresponding $\Lambda_{\mathrm{tot}}$
 are shown by the open circles.
 For GT transitions $L=0$, and the residual interaction with form
 factor (\ref{eq:r_L}) reduces to
 simple $(\vec{\sigma},\vec{\sigma})$-forces used in \cite{Gaarde1981}.
 This interaction cannot create high-excited $1^{+}$ states and cannot
 reproduce the effect of missing strength, as it was shown in \cite{Gaarde1981}.
 The experimental data extracted from
 \cite{SMR1987} are pointed by the crosses and the line presents
 the phenomenological estimations of \cite{GouPrim}
 with the parameters fitted in \cite{SMR1987}.
 In the calculation the transitions from
 the $J^{\pi}=0^{+}$ ground state to
 the final states with
 $J^{\pi} = 0^{\pm} , \ldots , 4^{\pm}$ were taken into account.
 The transitions to the states with higher momenta give negligible
 contributions to $\Lambda_{\mathrm{tot}}$.

\begin{figure}
 \includegraphics[width=0.75\linewidth]{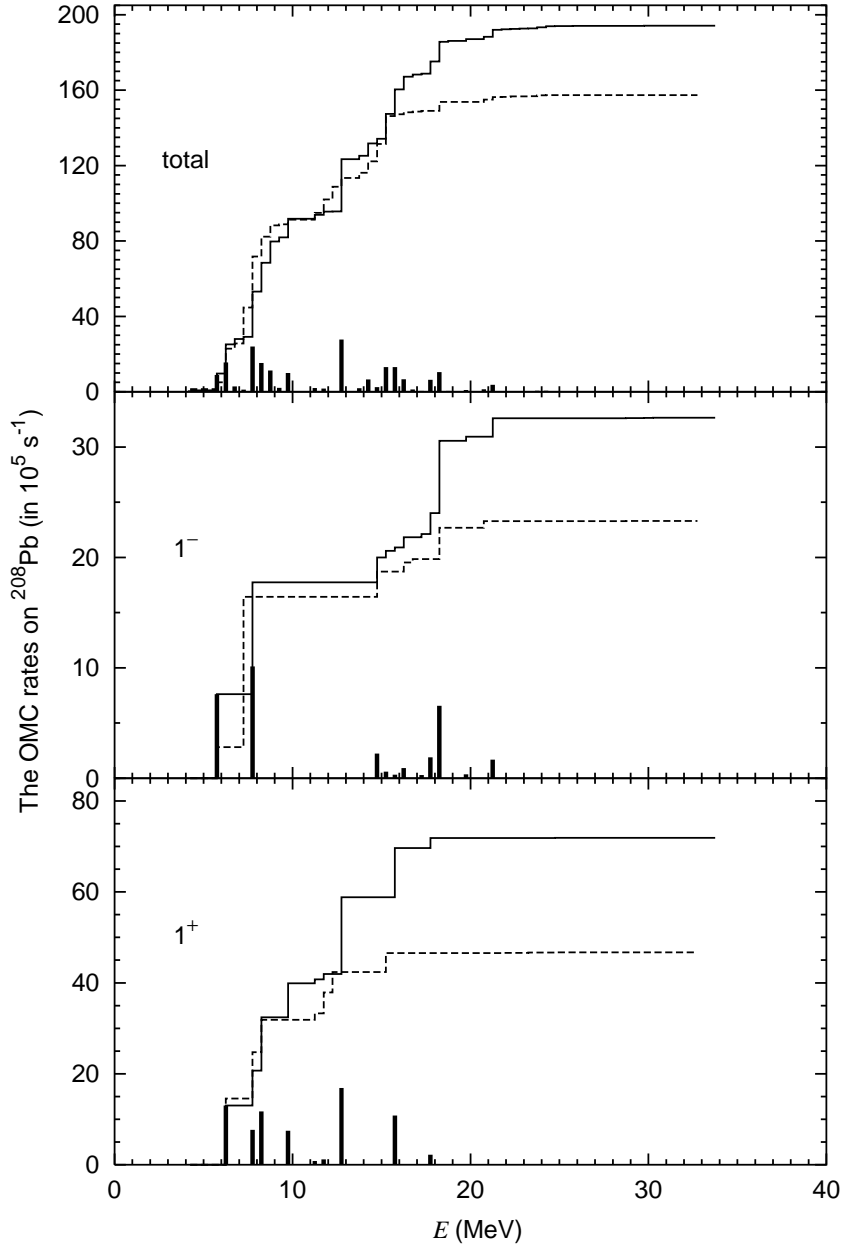}
 \caption{OMC rates by $\nucl{208}{Pb}$.
  Total and partial integrated capture rates up to the
  excitation energy $E$ of the final nucleus.
  The partial rates are shown for the final states
  with $J^{\pi} = 1^{\pm}$.
  Solid lines show the results of calculation with the radial
  form factor (\protect\ref{eq:dUdr}) of residual interaction,
  dashed lines show the same with the form factor
  (\protect\ref{eq:r_L}).
  The solid vertical bars present the distribution of
  calculated partial rates over the excited states with
  $J^{\pi} = 1^{\pm}$.
  In the upper part of the figure the distribution over
  all excited states is displayed.}
 \label{fig:pb208_OMC}
\end{figure}

\begin{table}
\caption{Total rates of OMC and relative contribution
 to $\Lambda_{\mathrm{tot}}$ of transitions to $1^{+}$ states.}
\label{tab:omc_heavy}
$$
  \begin{tabular}{ | c | c c | c c | c c | c | }
 \hline
  Target &
 \multicolumn{6}{c |}{%
 \begin{tabular}{@{} c @{} }
  Theoretical results for residual interaction \\
  with radial form factor \\
 \end{tabular}
  } &
 \begin{tabular}{@{} c @{} }
 Exper. \\
   $\Lambda_{\mathrm{tot}}$ \\
 \end{tabular} \\
 \cline{2-7}
 nucleus &
   \multicolumn{2}{c |}{No interaction} &
   \multicolumn{2}{c |}{ (\ref{eq:dUdr}) } &
   \multicolumn{2}{c |}{ (\ref{eq:r_L})} &
 (in $10^{5}\,\mathrm{s}^{-1}$) \\
 {} &
 $\Lambda_{\mathrm{tot}}$ & $\%$ of $1^{+}$ &
 $\Lambda_{\mathrm{tot}}$ & $\%$ of $1^{+}$ &
 $\Lambda_{\mathrm{tot}}$ & $\%$ of $1^{+}$ &
  \cite{SMR1987} \\
 \hline
  $\nucl{58}{Ni}$  &  $95.7$  &  $19$  &  $57.8$ & $28$ &  $63.0$ & $23$ &  $61.10 \pm 1.05$ \\
  $\nucl{60}{Ni}$  &  $88.2$  &  $18$  &  $46.7$ & $30$ &  $57.5$ & $22$ &  $55.62 \pm 0.97$ \\
  $\nucl{62}{Ni}$  &  $80.3$  &  $18$  &  $42.4$ & $30$ &  $52.3$ & $21$ &  $47.16 \pm 0.95$ \\
  $\nucl{90}{Zr}$  & $128.1$  &  $19$  &  $84.2$ & $35$ &  $87.4$ & $27$ &  $86.6 \pm 0.8$   \\
  $\nucl{92}{Mo}$  & $148.7$  &  $18$  &  $99.2$ & $31$ & $106.2$ & $25$ &  $96.2 \pm 1.5$   \\
 $\nucl{140}{Ce}$  & $208.4$  &  $21$  & $161.2$ & $36$ & $148.9$ & $24$ & $116.0 \pm 1.4$   \\
 $\nucl{208}{Pb}$  & $210.6$  &  $24$  & $197.2$ & $37$ & $159.5$ & $30$ & $134.5 \pm 1.8$   \\
 \hline
\end{tabular}
$$
\end{table}
 Figure\,\ref{fig:omc_tot} and Table\,\ref{tab:omc_heavy}
 show that for medium-weight nuclei
 (from nickel to tin) the theoretical $\Lambda_{\mathrm{tot}}$,
 calculated with both residual interactions
 agree with each other
 and reproduce correctly the experimental rates.
 Table\,\ref{tab:omc_heavy} shows that relative contributions to
 $\Lambda_{\mathrm{tot}}$ from transitions to the $1^{+}$ final states
 grow as the atomic number increases.
 For heavier nuclei ($\nucl{140}{Ce}$ and $\nucl{208}{Pb}$)
 theoretical $\Lambda_{\mathrm{tot}}$'s overestimate the experimental
 data considerably.
 The rates calculated with the form factor (\ref{eq:r_L}) are smaller
 than those obtained with the radial part (\ref{eq:dUdr}).
 Figure\,\ref{fig:pb208_OMC} shows that the difference comes
 about mainly due to capture feeding the high-excited $1^{\pm}$ states
 which are absent in calculations with $f_{LJ}(r) = r^{L}$.
 Table\,\ref{tab:omc_heavy}  shows that
 the excitation of high-excited states diminishes the
 influence of neutron excess on the muon capture.
 In rather heavy nuclei
 the muon capture rates
 (summed over all $0^{+} \rightarrow 1^{+}$ nuclear transitions)
 in calculation with the form factor (\ref{eq:dUdr})
 are larger than the rates calculated with the form factor (\ref{eq:r_L}),
 and even higher than the corresponding rates obtained in the model
 of independent quasiparticles (without any residual interaction).
 The fact that in heavier nuclei theoretical $\Lambda_{\mathrm{tot}}$'s
 overestimate the experimental rates was found in \cite{KLV2000}, where
 the Landau-Migdal residual forces were used.
 Their theoretical
 $\Lambda_{\mathrm{tot}}(\nucl{208}{Pb}) =
 161 \times 10^{5} \,\mathrm{s}^{-1}$
 and are well compared with the value
 $ 160 \times 10^{5} \,\mathrm{s}^{-1} $
 obtained in calculation with the residual interaction (\ref{eq:r_L}).

 The calculated $\Lambda_{\mathrm{tot}}$'s in heavy nuclei are found to
 be sensitive to the kind of residual interaction used.
 The largest discrepancy between the theory and experiment is obtained
 in the calculations with the residual interaction
 form factor (\ref{eq:dUdr}).
 This interaction forms the high-excited final states.
 As it was shown above, the high-excited $1^{+}$ states are responsible
 for the effect of missing GT strength.
 So the purpose of reproducing $\Lambda_{\mathrm{tot}}$ in heavy nuclei
 contradicts the description of the missing GT strength.

\section{Isovector $0^{+} \rightarrow 1^{+}$ transitions in $A=28$ nuclei}

\begin{figure}
\begin{minipage}[t]{0.48\linewidth}
 \includegraphics[bb=75 40 575 400,width=\linewidth]{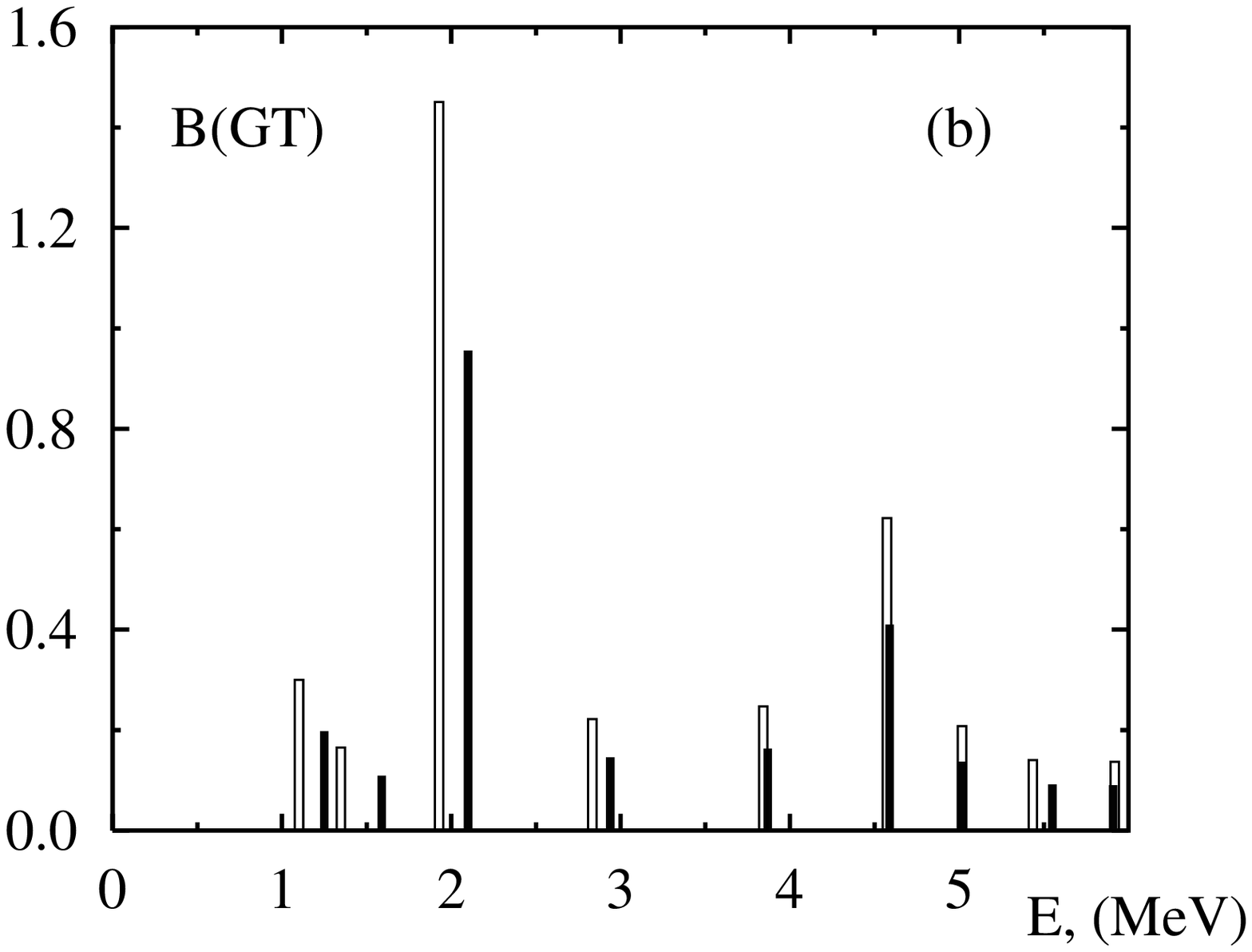}
\end{minipage}%
\begin{minipage}[t]{0.48\linewidth}
 \includegraphics[bb=75 40 575 400,width=\linewidth]{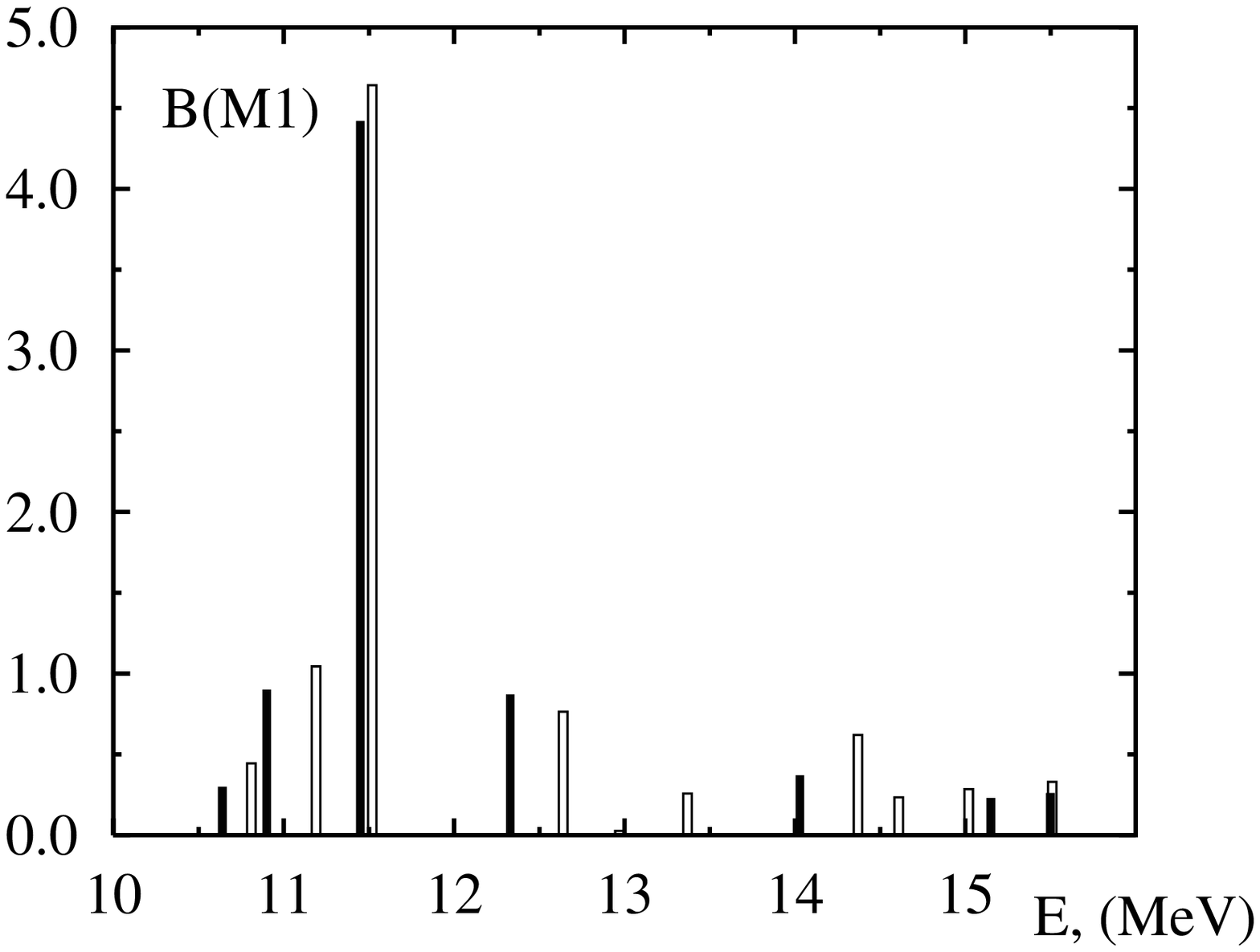}
\end{minipage}
\caption{Properties of isovector transitions
  in $A=28$ nuclei calculated with transformed wave functions.}
\label{fig:a28transf}
\end{figure}

 Unfortunately, it is impossible to extract from the experimental
 $\Lambda_{\mathrm{tot}}$ the capture rate determined by the
 $0^{+} \rightarrow 1^{+}$ nuclear transitions.
 One should look for partial muon capture in which the final state
 of a product nucleus is known.
 The experimental study of partial muon capture by $sd$-shell nuclei
 was carried out in \cite{omc_sd}.
 Here we discuss the transitions observed in the reaction
 \begin{equation}
 \label{eq:si28_omc}
   \mu^{-} + \nucl{28}{Si}(0^{+}_{g.s}) \rightarrow
    \nu_{\mu} + \nucl{28}{Al}(1^{+}) .
 \end{equation}
 The capture rates for transitions into three $1^{+}$ states
 with energies $1.620$, $2.201$ and $3.109$ MeV were measured in
 \cite{omc_sd}.
 These three states together with the $1^{+}$ states with energies
 $10.90$, $11.45$ and $12.33$ MeV in $\nucl{28}{Si}$
 and $1^{+}$ states with energies
 $1.59$, $2.10$, and $ 2.94$ MeV in $\nucl{28}{P}$
 form three isospin triplets \cite{Endt1990}.
 Starting from the $J^{\pi},T = 0^{+},0$ ground state of $\nucl{28}{Si}$
 the states of isospin triplets can be populated by spin-isospin probes:
 the states in $\nucl{28}{Al}$ --- by reaction (\ref{eq:si28_omc}).
 Equation (\ref{eq:partial_rate}) shows that the main part of
 the transitional operator is proportional to $\vec{\sigma} t^{+}$.
 The states in $\nucl{28}{Si}$ can be excited in the reaction of
 electron scattering $(e,e^{\prime})$; the transitional operator
 is proportional to $(g_{s} \vec{\sigma} + g_{l} \vec{l}\;) t^{0} $.
 The corresponding experimental data are published in \cite{si28_M1}.
 The states in $\nucl{28}{P}$ can be observed in
 $(p,n)$ reaction,
 in which the transition operator is proportional to $\vec{\sigma} t^{-} $;
 the experimental results are given in \cite{si28_GT}.
\begin{figure}
\begin{minipage}[t]{0.48\linewidth}
 \includegraphics[bb=75 40 575 400,width=\linewidth]{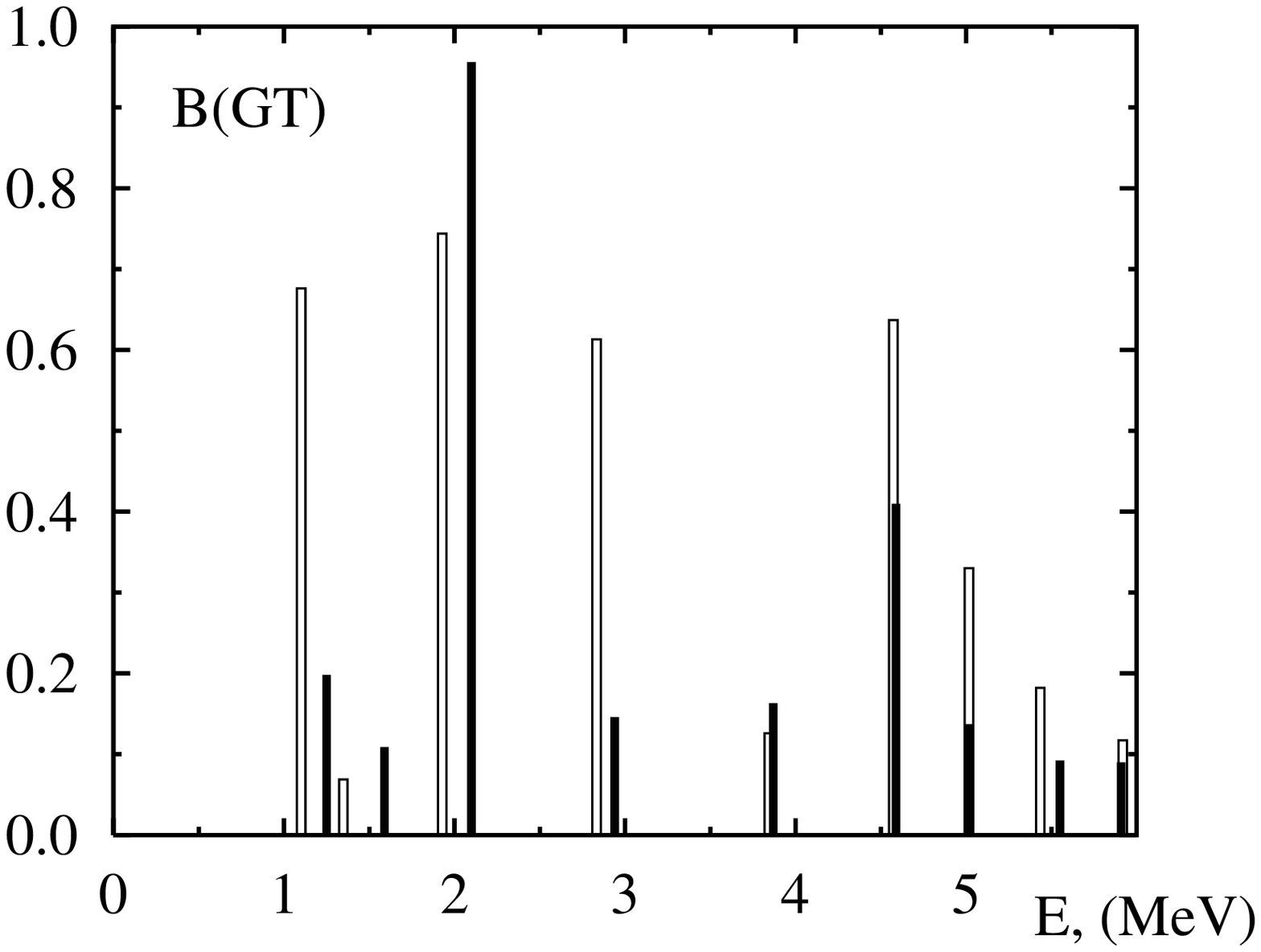}
\end{minipage}
\begin{minipage}[t]{0.48\linewidth}
 \includegraphics[bb=75 40 575 400,width=\linewidth]{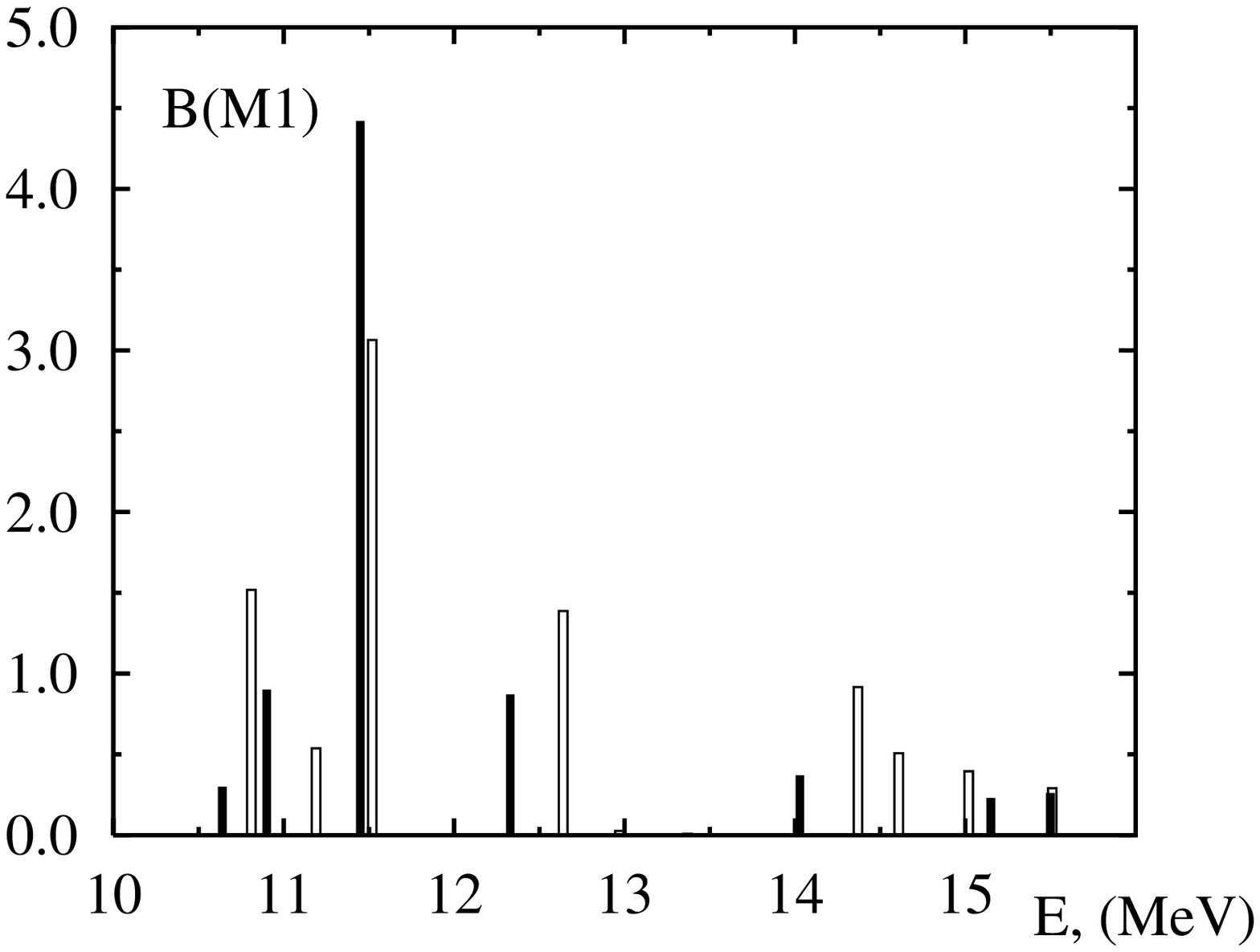}
\end{minipage}
\caption{Isovector $0^{+} \rightarrow 1^{+}$ transitions in $A=28$ nuclei.
 The experimental data are presented by closed bars.
 Open bars are theoretical results calculated
 with Hamiltonian \protect\cite{Wildenthal}. }
\label{fig:a28isov}
\end{figure}
 The experimental values of reduced probabilities of
 magnetic dipole, $B(M1)$, and GT transitions
 as functions of excitation energy are presented
 by the closed bars in Fig.\,\ref{fig:a28isov}.
 The open bars show the results of calculations within
 the multiparticle shell model with Wildenthal's
 Hamiltonian \cite{Wildenthal}.
 The computer code OXBASH \cite{OXBASH} was used in calculations.
 The $B(M1)$ values were obtained with free $g_{s}$ and $g_{l}$ factors,
 and no effective charges were used in $B(GT)$ calculations.
 In principle, one should expect introducing the effective charge
 because the shell model \cite{Wildenthal} works within
 the $sd$-shell space,
 and the previous discussion shows the importance of $\Delta N \geq 2$
 transitions for correct description of the GT strength function, at least.
 In the present situation it is a difficult task to determine the effective
 charge because theoretical $B(M1)$ and $B(GT)$
 values are higher than experimental ones for many states,
 except the strongest transition in which the theory goes below experiment
 for both $(p,n)$ and $(e,e^{\prime})$ reactions.
 Theoretical summed transition strengths are higher than experimental
 ones:
 $B_{\Sigma}^{\mathrm{exp}}(GT) = 0.66 \, B_{\Sigma}^{\mathrm{th}}(GT)$
 and
 $B_{\Sigma}^{\mathrm{exp}}(M1) = 0.85 \, B_{\Sigma}^{\mathrm{th}}(M1)$.
 To cope with this situation, an orthogonal transformation acting in subspace
 of wave functions of isovector $1^{+}$ states was suggested in
 \cite{KT2000}.
 The parameters of the transformation were chosen such that the theoretical
 strength functions of GT and $M1$ transitions calculated with transformed
 wave functions coincided in shape (within a constant factor) with
 the corresponding experimental functions.
 Due to orthogonality of transformation, theoretical summed
 transition strengths are conserved, and the transformation results in
 redistribution of $B(GT)$'s and $B(M1)$'s over excitation energy.
 The transformed strength functions are presented
 in Fig.\,\ref{fig:a28transf}.
 The transformation established exact proportionality between
 theoretical and experimental GT strength functions and
 approximate proportionality for $M1$ strength functions.

\begin{table}
\caption{Properties of isovector $1^{+}$ states
 in $A=28$ nuclei;
 $\tau$-- life-time;
 $\gamma$-- $\gamma$-decay branching ratios;
  $\Lambda_{f}$ -- partial OMC rate.}
\label{tab:a28}
\begin{center}
\begin{tabular}{| c | l | c c | c | c | }
 \hline
 Nucleus & \multicolumn{1}{c |}{Observable}  &
 \multicolumn{2}{c|}{ Experiment} &
 \multicolumn{2}{c|}{ Theory } \\
 \cline{3-6}
  {}  & {}    & Value & Ref. &  (a) &  (b) \\
 \hline
 $\nucl{28}{Al}$ &
 $\tau(1^{+}_{2.201})$ (in fs) &
 \begin{tabular}{@{} c @{}}
 $ 59 \pm 6 $ \\
 $ 51 \pm 1.0_{\mathrm{stat.}} \pm 6.7_{\mathrm{sys.}} $ \\
 \end{tabular}
 &
 \begin{tabular}{@{} c @{}}
 \cite{Endt1998}      \\
 \cite{Briancon2000}  \\
 \end{tabular}
 &
 66 & 49 \\
 {} &
 $ \gamma(1^{+}_{2.201} \to 2^{+}_{0.031}) $ (in \%) &
 $ 79  $ & \cite{Endt1990} &
 $  2  $ & $ 63 $ \\
 {} &
 $ \gamma(1^{+}_{2.201} \to 0^{+}_{0.972}) $ (in \%) &
 $ 16  $ & {} &
 $ 80  $ & $ 35 $ \\
\hline
 $\nucl{28}{Al}$ &
 $\Lambda(1^{+}_{1.620})$ (in $10^{3} \, \mathrm{s}^{-1}$) &
 $ 12.9 \pm 2.1 $ & \cite{omc_sd} &
 $ 3.1 $ & $ 7.6 $ \\
 {} &
 $\Lambda(1^{+}_{2.201})$ (in $10^{3} \, \mathrm{s}^{-1}$) &
 $ 62.8 \pm 7.4 $ & {} &
 $ 34.1 $ & $ 63.6 $ \\
 {} &
 $\Lambda(1^{+}_{3.110})$ (in $10^{3} \, \mathrm{s}^{-1}$) &
 $ 14.7 \pm 2.6 $ & {} &
 $ 26.1 $ & $ 11.2 $ \\
\hline
 $\nucl{28}{Si}$ &
 $ B_{M1}(1^{+}_{10.90}) $ ($\mu_{N}$) &
 $ 0.90 \pm 0.02 $ & \cite{si28_M1} &
 $ 0.54 $ & $ 1.04 $ \\
 {} &
 $ B_{M1}(1^{+}_{11.45}) $ ($\mu_{N}$) &
 $ 4.42 \pm 0.20 $ & {} &
 $ 3.06 $ & $ 4.46 $ \\
 {} &
 $ B_{M1}(1^{+}_{12.33}) $ ($\mu_{N}$) &
 $ 0.87 \pm 0.06 $ & {} &
 $ 1.39 $ & $ 0.76 $ \\
\hline
 $\nucl{28}{P}$ &
 $ B_{GT}^{-}(1^{+}_{1.59}) $ &
 $ 0.109 \pm 0.002 $ & \cite{si28_GT} &
 $ 0.069 $ & $ 0.165 $ \\
 {} &
 $ B_{GT}^{-}(1^{+}_{2.10}) $ &
 $ 0.956 \pm 0.005 $ & {} &
 $ 0.774 $ & $ 1.451 $ \\
 {} &
 $ B_{GT}^{-}(1^{+}_{2.94}) $ &
 $ 0.146 \pm 0.003 $ & {} &
 $ 0.613 $ & $ 0.222 $ \\
\hline
\end{tabular}
\end{center}
\end{table}

 The characteristics of partial transitions to the members
 of isospin triplets are collected in Table\,\ref{tab:a28}.
 Theoretical values calculated with the initial wave functions
 (column (a)) and with transformed wave functions (column (b))
 are presented in Table\,\ref{tab:a28} for comparison.
 The muon capture rates were calculated with free values of
 weak nucleon current couplings.
 The transformation of the wave functions of excited states improves
 greatly the description of muon capture rates, and for the strongest
 transition the theoretical rate coincides with experiment.
 The calculation with the transformed wave functions reproduces
 correctly the experimental $B(M1)$ values.
 However, there is a considerable disagreement between theoretical
 (b) and experimental $B(GT)$'s.
 It is an unexpected result because the spin-isospin parts
 of the operators describing the charge-exchange reactions,
 magnetic scattering of electrons, and muon capture
 are practically the same.
 The discrepancy between theoretical and experimental
 $B(GT)$ values indicates that the relation between the cross sections
 of charge-exchange reactions and $B(GT)$'s may be complicated
 even for strong transitions.

\section{Conclusion}

 The energy weighted moments of strength
 function of GT transition, $S^{\pm}_{k}$, are calculated
 in the RPA, SRPA and within the fragmentation problem.
 Considering $S^{-}_{0}$ and $S^{-}_{1}$ we have shown
 that the effect of missing GT strength should be reproduced
 as the result of interaction among the particle-hole excitations,
 without including the 2p--2h configurations.
 Hence, the residual interaction in the spin-isospin
 channel must intensively mix the $\Delta N =0$ and $\Delta N \geq 2$
 particle-hole states.
 The example of this interaction is presented.
 It is shown that the experimental strength function
 of $\sigma t^{-}$ transition in $\nucl{90}{Zr}$ can be
 reproduced rather well in the whole region of excitation energy.

 Total muon capture rates were calculated for several nuclei
 using two variants of residual interaction.
 Theoretical total rates of muon capture by medium nuclei practically
 do not depend on the residual interaction used in calculation.
 In heavy nuclei theoretical rates are higher than
 experimental ones.
 The excess depends on the residual interaction, and the difference
 between theory and experiment is the largest when the residual
 interaction, which forms the high-excited collective states,
 is used in the calculation.
 The existence of these states is assumed by the effect of missing
 GT strength.

 It is shown that the distributions of transition strength over
 the excitation energy in $A=28$ nuclei extracted from weak and
 electromagnetic processes are in conflict with the ones obtained
 from charge-exchange nuclear reactions.
 In particular, no quenching of spin-isospin transitions is found
 in the rates of partial allowed muon capture
 $\nucl{28}{Si}(0^{+}_{g.s})(\mu, \nu) \nucl{28}{Al}(1^{+})$.


\begin{thebibliography}{ii}
%
\bibitem{Rowe68} %
 D.\,J.\ Rowe, Rev.\ Mod.\ Phys., {\bf 40,} 153, (1968)
%
\bibitem{LaneMart80} %
 A.\,M.\ Lane and J.\ Martorell, Ann.\ Phys., {\bf 129,} 273, (1980)
%
\bibitem{Gantmacher}
 F.\ Gantmacher, Theory of Matrices, vol.\ 2, AMS Publishing, 2000
%
\bibitem{DNSW90} %
 S.\ Dro\.zd\.z, S.\ Nishizaki, J.\ Speth and J.\ Wambach,
 Phys.\ Rep., {\bf 197,} 1, (1990)
%
\bibitem{BM} %
 A.\ Bohr and B.\,R.\ Mottelson,
 Nuclear Structure, vol.\ 1,
 New York, Amsterdam, 1969
%
\bibitem{VS1983} %
 V.\,V.\ Voronov and V.\,G.\ Soloviev,
 Fiz.\ Elem.\ Chast.\ At.\ Yadra, {\bf 14,} 1380, (1983)
 [Sov.\ J.\ Part.\ Nucl., {\bf 14,} 583 (1983)]
%
\bibitem{KuzFrag} %
 V.\,A.\ Kuz'min, Theoret.\ i Matemat.\ Fiz., {\bf 70,} 315 (1987)
 [Theor.\ and Math.\ Phys., {\bf 70,} 223 (1987)]
%
\bibitem{Bert83} %
 G.\,F.\ Bertsch, P.\, F.\ Bortignon and R.\,A.\ Broglia,
 Rev.\ Mod.\ Phys., {\bf 55,} 287 (1983)
%
\bibitem{Mat83} %
 G.\,J.\ Mathews, S.\,D.\ Bloom and R.\,F.\ Hausman, Jr.,
 Phys.\ Rev.\ C, {\bf 28,} 1367 (1983)
%
\bibitem{DKSW86}
 S.\ Dro\.zd\.z, V.\ Klempt, J.\ Speth, and J.\ Wambach,
 Phys.\ Lett.\ B, {\bf 166,} 18 (1986)
%
\bibitem{DO1987}
 S.\ Dro\.zd\.z, F.\ Osterfeld, J.\ Speth, J.\ Wambach,
 Phys.\ Lett.\ B, {\bf 189,} 271 (1987)
%
\bibitem{VdoSol1983} %
 A.\,I.\ Vdovin and V.\,G.\ Soloviev,
 Fiz.\ Elem.\ Chast.\ At.\ Yadra, {\bf 14,} 237, (1983)
 [Sov.\ J.\ Part.\ Nucl., {\bf 14,} 99 (1983)]
%
\bibitem{Wakasa1997} %
 T.\ Wakasa {\it et al.}, Phys.\ Rev.\ C, {\bf 56,} 2909, (1997)
%
\bibitem{examples} %
 V.\,A.\ Kuz'min, Yad.\ Fiz.,  {\bf 58,} 418, (1995)
 [Phys.\ Atom.\ Nucl.. {\bf 58,} 368, (1995)];
 K.\ Junker, V.\,A.\ Kuz'min, T.\,V.\ Tetereva,
 Eur.\ Phys.\ J.\ A, {\bf 5,} 37 (1999)
%
\bibitem{BKE1978} %
 V.\,V.\ Balashov, G.\,Ya.\ Korenman, R.\,A.\ Eramzhyan,
 Poglozhenie mezonov atomnymi yadrami,
 M., Atomizdat, 1978, 294 pp (in Russian)
%
\bibitem{EKTJO} %
 R.\,A.\ Eramzhyan, V.\,A.\ Kuz'min, and T.\,V.\ Tetereva,
 Nucl.\ Phys.\ A, {\bf 642,} 428 (1998);
 V.\,A.\ Kuzmin, T.\,V.\ Tetereva, K.\ Junker, and
 A.\,A.\ Ovchinnikova,
 J.\ Phys.\ G: Nucl.\ Part.\ Phys., {\bf 28}, 665 (2002)
%
\bibitem{Gaarde1981} %
 C.\ Gaarde {\it et al.,} Nucl.\ Phys.\ A, {\bf 369,} 258 (1981)
%
\bibitem{SMR1987} %
 T.\ Suzuki, D.\,F.\ Measday, and J.\,P.\ Roalsvig,
 Phys.\ Rev.\ C, {\bf 35,} 2212 (1987)
%
\bibitem{GouPrim} %
 B.\ Goulard and H.\ Primakoff,
 Phys.\ Rev.\ C, {\bf 10,} 2034 (1974)
%
\bibitem{KLV2000} %
 E.\ Kolbe, K.\ Langanke, and P.\ Vogel,
 Phys.\ Rev.\ C, {\bf 62}, 055502 (2000)
%
\bibitem{omc_sd} %
 T.\,P.\ Gorringe, {\it et al.,}
 Phys.\ Rev.\ C, {\bf 60,} 055501 (1999)
%
\bibitem{Endt1990} %
 P.\,M.\ Endt,
 Nucl.\ Phys.\ A, {\bf 521}, 1 (1990)
%
\bibitem{si28_M1} %
 C.\ L\"uttge, {\it et al.,}
 Phys.\ Rev.\ C, {\bf 53,} 127 (1996);
 Y.\ Fujita, {\it et al.,}
 Phys.\ Rev.\ C, {\bf 55,} 1137 (1996)
%
\bibitem{si28_GT} %
 P.\ von Neumann-Cosel,
 A.\ Richter, Y.\ Fujita and B.\,D.\ Anderson,
 Phys.\ Rev.\ C, {\bf 55,} 532 (1997)
%
\bibitem{Wildenthal} %
 B.\,H.\ Wildenthal, Prog.\ Part.\ Nucl.\ Phys.,
 {\bf 11,} 5 (1984)
%
\bibitem{OXBASH} %
 A.\ Etchegoyen., B.\,A.\ Brown, W.\,D.\,M.\ Rae,
 MSUCL Report No.\ 524, Michigan, 1986
%
\bibitem{KT2000} %
 V.\,A.\ Kuz'min and T.\,V.\ Tetereva,
 Yad.\ Fiz., {\bf 63,} 1966 (2000)
 [Phys.\ At.\ Nucl., {\bf 63,} 1874 (2000)]
%
\bibitem{Endt1998} %
 P.\,M.\ Endt,
 Nucl.\ Phys.\ A, {\bf 633}, 1 (1998)
%
\bibitem{Briancon2000} %
 Ch.\ Brian\c{c}on, {\it et al.,}
 Nucl.\ Phys.\ A, {\bf 671}, 647 (2000)
%
\end{thebibliography}
\end{document}